\documentclass[sigconf]{acmart}

\usepackage{booktabs}
\usepackage{tabularx}
\usepackage{enumitem}
\usepackage{xcolor}
\usepackage{url}
\usepackage{listings}
\usepackage{microtype}
\usepackage{makecell}
\newif\ifshowannotations
\showannotationstrue  

\title{From Threads to Trajectories: A Multi-LLM Pipeline for Community Knowledge Extraction from GitHub Issue Discussions}

\author{Nazia Shehnaz Joynab}
\email{nazia.joynab@utdallas.edu}
\affiliation{%
  \institution{The University of Texas at Dallas}
  \country{USA}
}

\author{Soneya Binta Hossain}
\email{sbhossain@utdallas.edu}
\affiliation{%
  \institution{The University of Texas at Dallas}
  \country{USA}
}

\begin{document}

\begin{abstract}
Resolution of complex post-production issues in large-scale open-source software (OSS) projects requires significant cognitive effort, as developers need to go through long, unstructured and fragmented issue discussion threads before that. In this paper, we present \textbf{SWE-MIMIC-Bench}, an issue trajectory dataset generated from raw GitHub discussions using an automated multi-LLM pipeline. Unlike simple summarization, this pipeline utilizes a group of closed-source LLMs to perform granular tasks: analyzing individual comments with awareness of externally-linked resources, classifying comment analyses into label-specific fields (e.g., root cause, solution plan, implementation progress), and synthesizing label-aware trajectories which capture a structured and coherent narrative of the entire discussion thread. Our pipeline uses five closed-source LLM configurations for distinct purposes: label classification, inline code block and external link summarization, comment analysis, label-specific field classification and trajectory synthesis. By generating concise and reliable trajectories from complex conversation threads, this system can assist developers and researchers of broader software engineering community to understand the experience-driven collaborative approach for issue diagnosis. Furthermore, the generated trajectories can be used to train modern LLM agents to think and act like an expert developer. We evaluated our system on 800 real-world GitHub issues drawn from the SWE-Bench-Pro, SWE-Bench-Multilingual and SWE-Bench-Verified dataset, achieving a 91.7\% success rate in extracting 734 high-fidelity reasoning trajectories.
\end{abstract}

\maketitle

\section{Introduction}
\label{sec:intro}

In large-scale open-source software (OSS) development, issues are critically diverse and require continuous community discussions over months or even years. For bugs, it is needed to determine the underlying root cause before implementing and deploying a patch, whereas for feature-requests, it is required to assess if the planned changes will improve the system’s reliability or negatively impact other working components. Therefore, the long discussion threads on varied software issues can create a valuable resource of knowledge on varied issue-comprehension processes. Understanding the context of these threads becomes exemplary, when faced with similar issues from the past. However, it takes a significant amount of time and effort, which leads to reduced productivity of a developer. Even, for an autonomous agent which attempts to replicate expert-level repair, this community knowledge can be very important to avoid wrong trajectories or reasoning pathways which were already explored by the human developers.

Recent advancements in agentic AI have made substantial progress in automating end-to-end software engineering tasks \cite{yang2024swe}. Given a problem description and access to the codebase of an OSS project, an agent can generate candidate code patches and validate them against given test suites. Current benchmarks such as SWE-Bench \cite{jimenez2023swe}, SWE-Bench-Pro \cite{deng2025swe} are playing a key role in evaluating this capability. Modern agentic systems can also access community discourse on public platforms like Jira tickets or Github issue discussion threads, yet their reasoning processes do not explicitly mention how they are harnessing the extracted knowledge from these threads. The main insights of the issue resolution process including issue diagnosis, root cause analysis, rejected hypothesis, final decisions etc. are often distributed across multiple issue threads in a very unstructured way. Without the entire context of all the relevant threads of a complex issue, solely the issue-header and problem description can be insufficient for an agent to resolve it in an efficient manner. 

Recently, researchers are starting to address this gap by modeling intermediate reasoning traces over community discussion threads, e.g., SVRC (Structured and
Validated Reasoning Chains for Code Generation) \cite{yang2025think}. While the research represents a meaningful starting contribution on this track, its methodology is optimized for coding problems rather than real-world software engineering tasks. In most of the well-maintained software systems, issues can be categorically diverse. A reasoning trajectory representing a crash report builds upon problem description, issue reproduction steps, temporary mitigation plans, affected components, stack traces etc. and therefore it will not be well-suited for a feature request, which demands feature description, use-case analysis, design decisions and implementation progress. So, applying a monolithic template across categorically different issues can result in misleading comprehension of those issues. Consequently, this can prevent a model from learning how to adapt their thinking paths efficiently when faced with any unseen issue.

\textbf{Why existing approaches are not enough.}
There are three main reasons why current research is unable to adequately capture the process-level reasoning present in discussions of actual software issues.
\textit{First}, in SWE-Bench and its descendant benchmarks, issue resolution is treated as a static mapping from a problem description to a final resolution. The complex, temporal evolution of reasoning seen in issue threads, such as iterative hypotheses, debugging dead-ends, and cooperative consensus, is completely ignored in these benchmarks. As a result, they fail to provide a real exposure of the cognitive workflow of developers.
\textit{Second}, whereas newer initiatives such as SWE-Synth \cite{pham2025swe} synthesize intermediate repair traces in an attempt to mimic step-by-step debugging strategy of developers, these workflows are not based on actual community dialog. Most importantly, they are not label-guided. The lack of adaptive semantic structures in their intermediary processes (e.g., differentiating between use-case planning for features and root cause analysis for bugs) limits their downstream effectiveness and interpretability.
\textit{Third}, a strict, pre-established Software Development Life Cycle (SDLC) pattern is enforced by reasoning-oriented datasets such as SVRC \cite{yang2025think}. SVRC presupposes a standardized workflow that ignores the varied, context-dependent, and artifact-rich nature of actual open-source maintenance as it is built from curated coding platforms rather than natural GitHub discussions.
In the end, none of these methods result in label-guided trajectories that maintain the collaborative aspect of issue discussions. Because of this gap, neither autonomous agents nor software enthusiasts get access to the structured, human-grounded reasoning data required to mimic expert-level software engineering.

\textbf{Our goal.}
Our main goal is to make the community knowledge obtained from GitHub issue threads more explicit, organized, and machine-accessible in order to improve the reasoning processes of automated software engineering agents. Rather than using rigid reasoning templates or straightforward problem-to-patch translations, our work methodically recovers label-guided trajectories from the raw discourse of GitHub issue threads. In order to capture the unique cognitive workflow involved in issue comprehension, our approach dynamically classifies the extracted reasoning texts into fields that belong to the particular category or label of the issue. To ensure broader domain range for the extracted knowledge, we developed our dataset by aggregating issues from multiple variants of SWE-Bench, such as SWE-Bench-Pro \cite{deng2025swe}, SWE-Bench-Multilingual \cite{jimenez2023swe} and SWE-Bench-Verified \cite{jimenez2023swe}. Because of the diversified issue types, complexities, and conversation patterns in these three datasets, the gathered trajectories by our methodology represent a variety of reasoning processes. As a result, these trajectories can offer deep insights into how issue comprehension processes evolve under various circumstances. 


\textbf{Our contributions.}
In this paper, we present a multi-LLM framework that automates the extraction of community knowledge from GitHub issue threads to support software issue diagnosis and resolution. Specifically, our work makes the following contributions:

\begin{enumerate}
\item \textbf{Label-Guided Trajectory Generation Method:} We introduce an automated method for extracting dynamic, label-aware reasoning trajectories from GitHub issue discussions. To the best of our knowledge, this is the first framework that explicitly models the diverse reasoning processes employed by expert developers across different categories of software tasks.

\item \textbf{Artifact-Grounded Discussion Analysis:} Our approach addresses the limitation of \emph{artifact blindness} in existing text-centric benchmarks by incorporating context from external resources such as code snippets, commit links, images, and documentation. These artifacts are pre-fetched, summarized, and integrated into comment-level analysis, enabling more comprehensive and reliable trajectory synthesis.

\item \textbf{A Foundational Dataset for Agentic RL:} We construct and release a novel dataset namely \textbf{SWE-MIMIC-BENCH}, consisting of 734 high-fidelity, structured reasoning trajectories which were transformed from real-world GitHub issue threads. This dataset provides a valuable resource for training next-generation autonomous software engineering agents, particularly in reinforcement learning settings for root cause analysis and automated program repair.
\end{enumerate}

\section{Dataset Overview}
\textbf{SWE-MIMIC-Bench} is an LLM-generated dataset consisting of 734 label-guided issue reasoning trajectories from GitHub. These issues are primarily derived from SWE-Bench-Pro (test set)~\cite{deng2025swe}, SWE-Bench-Multilingual (test set)~\cite{jimenez2023swe}, and SWE-Bench-Verified (test set)~\cite{jimenez2023swe} dataset.

Starting from an initial pool of approximately 800 issue threads, trajectories that failed to meet quality criteria, such as insufficient field coverage, lack of coherence, or factual inconsistencies, were systematically filtered out using manual inspection and LLM-based evaluation. As a result, the final dataset consists of 734 high-quality trajectories that preserve the essence of the original discussions.

\subsection{Dataset Unit}
Each unit in SWE-MIMIC-Bench is defined by a matched pair of files. The \textbf{input file}, named \textit{issue\#{N}\_comments\_pr\#{M}.json}, is a structured JSON export of a single GitHub issue thread, containing the issue metadata (source URL, repository owner's name, repository name, issue number, total comment count) and the full list of comments. Each comment record includes the comment ID, author login, type, association, bot flag, creation and update timestamp, body text, and a reactions dictionary. One comment is marked \textit{is\_header: true} to identify the original issue post. The \textbf{output file}, named \textit{{N}\_issue\_trajectory.json}, is the structured trajectory produced by our developed pipeline for that issue. It preserves the full reasoning chain from the entire thread, containing the issue title, labels, detection method of labels, label-specific field schema, a URL-to-summary link cache, per-comment free-form analyses, field-bucketed evidence collections, and the synthesized trajectory itself. 

\subsection{Issue Labels}
Our method utilizes six existing GitHub issue labels for final trajectory generation: \textit{bug, enhancement, help\_wanted, good\_first\_issue, documentation, and question}. These labels are supplemented by a generic fallback schema. Each label triggers a specialized \textit{Field Schema}, where each field in the schema is designed to extract the specific reasoning signals relevant to that field type. We refer this process as \textit{field-bucketed evidence collection.}

\subsection{Trajectory Field Schema}
As the developed pipeline generates label-guided trajectories, the trajectory schema is not uniform across all issues. Each issue label maps to a set of fields that capture the cognitive structure of that label. For \textbf{bug} type issues, the trajectory captures the following fields: problem description, reproduction steps, scope and impact, root cause analysis, workaround, solution plan, decision consensus, testing and verification. 
For \textbf{enhancement} type issues, the trajectory captures: feature description, motivation and use case, solution approaches, technical challenges, design decisions, and implementation progress. 
For \textbf{question} type issues, the schema narrows to: question asked, context and evidence, clarification requests, answers and explanations, root cause identified, and resolution outcome.
For \textbf{help\_wanted} issues, the fields are: task description, proposed solution, technical context, dependencies and scope, contributor engagement, and current status.
For \textbf{good\_first\_issue} threads, the schema emphasizes: task description, beginner-friendly rationale, technical requirements, implementation guidance, reference examples, mentor interaction, and progress tracking.
For \textbf{documentation} issues, the fields cover: documentation issue type, affected components, proposed changes, maintainer response, dependencies and blockers, and current status.
Issues whose labels do not match any predefined keywords are classified as \textbf{general}, which is a fall back to a five-field default schema: issue description, context and background, discussion points, proposed actions, and decision consensus. Fields for which no meaningful evidence is found across the thread are assigned a null value rather than a fabricated summary, preserving the fidelity of the extracted narrative.

\subsection{Summary Statistics}
Table \ref{tab:dataset_stats} reports the issue type distribution across the three SWE-Bench variants successfully processed by our automated method. Across all three splits, \textbf{bug} issues constitute the dominant category, accounting for 85.2\% of the 258 successfully processed trajectories in the Multilingual split, 84.4\% of the 263 in the Verified split, and 57.2\% of the 213 in the Pro split. The Pro split exhibits a notably more balanced distribution: \textbf{enhancement} issues represent 36.6\% of its trajectories, compared to only 9.7\% in Multilingual and 13.3\% in Verified, reflecting the broader diversity of issue types present in the Pro benchmark. Minority categories such as \textbf{good first issue}, \textbf{help wanted}, \textbf{question}, and \textbf{documentation} each account for fewer than 5\% of trajectories in any single split. No \textbf{general} issues appear in any split, indicating that the pipeline's two-stage classification: keyword matching followed by LLM fallback successfully assigned a recognized label type to every processed thread. These distributions confirm that our proposed label-guided schema is exercised across a realistic and varied mixture of issue categories.
\begin{table}[t]
\label{tab:dataset_stats}
\centering
\caption{Issue type distributions in SWE-MIMIC-Bench over successfully processed trajectories.}
\footnotesize
\setlength{\tabcolsep}{4pt}
\renewcommand{\arraystretch}{1.05}
\resizebox{\linewidth}{!}{%
\begin{tabular}{p{3.2cm}|rr|rr|rr}
\hline
& \multicolumn{2}{c|}{\textbf{Multilingual}} & \multicolumn{2}{c|}{\textbf{Verified}} & \multicolumn{2}{c}{\textbf{Pro}} \\
\textbf{Category} & \textbf{Count} & \textbf{\%} & \textbf{Count} & \textbf{\%} & \textbf{Count} & \textbf{\%} \\
\hline
\multicolumn{7}{l}{\textbf{(b) Issue Type Distribution}} \\
\hline
Bug               & 220 & 85.2 & 222 & 84.4 & 122 & 57.2 \\
Enhancement       &  25 & 9.7 &  35 & 13.3 &  78 & 36.6 \\
Good First Issue  &   4 &  1.6 &   0 &  0.0 &   6 &  2.8 \\
Help Wanted       &   8 &  3.1 &   1 &  0.4 &   1 &  0.4 \\
Question          &   1 &  0.4 &   3 &  1.1 &   4 &  1.9 \\
Documentation     &   0 &  0.0 &   2 &  0.8 &   2 &  0.9 \\
General           &   0 &  0.0 &   0 &  0.0 &   0 &  0.0 \\
\hline
\textbf{Total Approved} & \textbf{258} & \textbf{100.0} & \textbf{263} & \textbf{100.0} & \textbf{213} & \textbf{100.0} \\
\hline

\end{tabular}}
\label{tab:dataset_stats}
\end{table}

\section{Data Curation Pipeline}
\begin{figure*}[htbp]
    \centering
    \includegraphics[width=\linewidth]{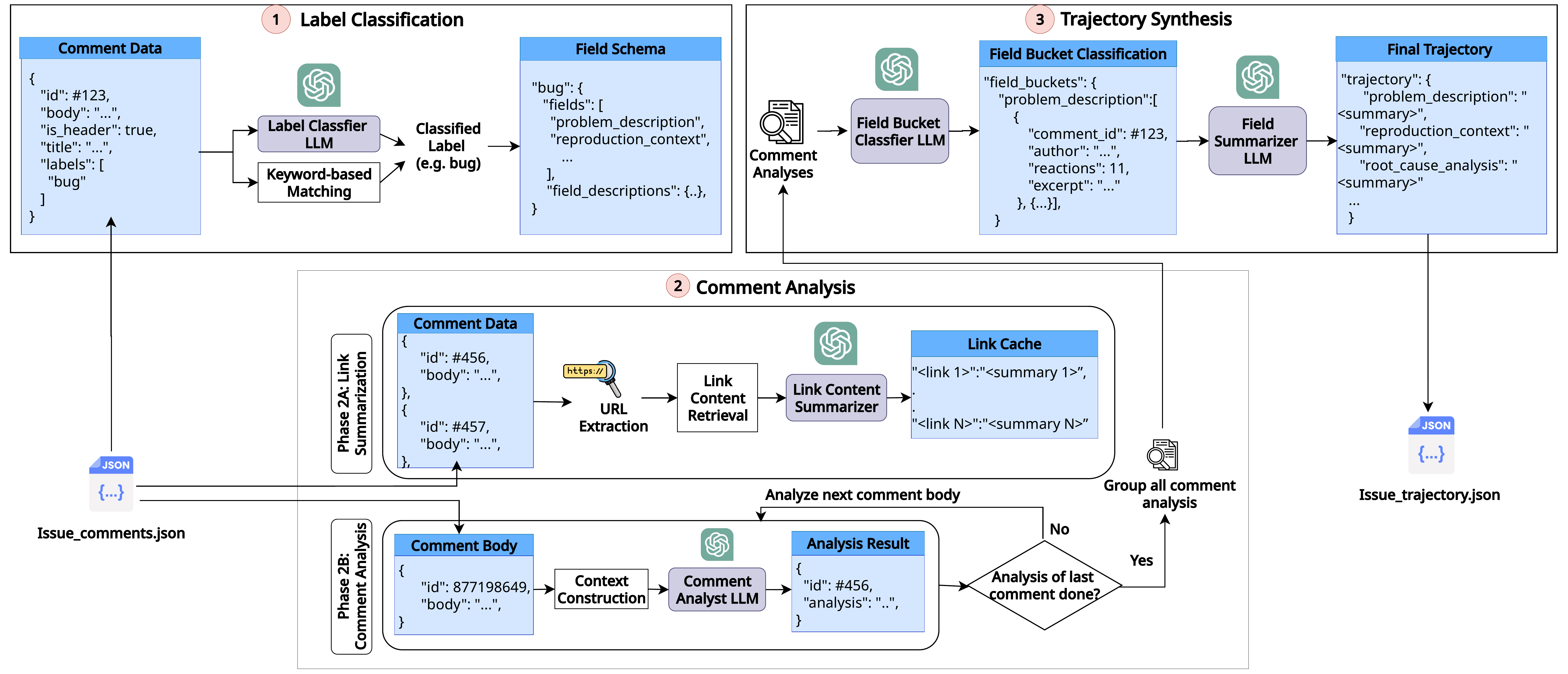}
    \caption{Overview of Issue Trajectory Generation Approach.} 
    \label{fig:approach}
\end{figure*}
We developed a sequential, multi-stage LLM pipeline (shown in Figure \ref{fig:approach}) to convert unstructured, raw GitHub issue threads into structured, cohesive narratives. This pipeline ensures the generation of reliable, artifact-grounded, and categorically appropriate reasoning trajectories by breaking down the difficult process of issue comprehension into specialized sub-tasks.
\subsection{LLM Configuration}
Rather than relying on a single monolithic prompt or model, our system utilizes an ensemble of specialized LLMs, with tasks dynamically routed based on their required reasoning depth and output constraints. This design enables a distinct separation between deterministic classification tasks and open-ended generative reasoning, improving both reliability and efficiency. To achieve this balance, we utilize two foundation models via the OpenAI API, configuring each component according to its functional role:
\begin{itemize}
\item \textbf{Deterministic Classifiers:} For tasks requiring strict, reproducible categorizations, specifically the \textit{Label Classifier LLM} and the \textit{Field Bucket Classifier LLM}, we deploy \textbf{gpt-4o-mini}. These models were configured with a temperature of $0.0$ to eliminate stochasticity and ensure consistent adherence to predefined schemas. Additionally, their generation horizons are tightly constrained (maximum 128 and 256 tokens, respectively) to optimize latency and cost.
\item \textbf{Generative Analyzers:} For tasks demanding deep contextual reasoning and technical summarization, namely the \textit{Comment Analyst LLM} and the \textit{Link Content Summarizer LLM}, we utilize the advanced \textbf{gpt-5.4-mini} model. These agents were configured with an extended output window of 4,096 tokens and a low but non-zero temperature ($0.2$), allowing for necessary linguistic flexibility while preventing factual hallucination.
\item \textbf{Trajectory Synthesizer:} Finally, to combine the categorized bucketed data into cohesive, paragraph-level narratives, we utilize \textbf{gpt-4o-mini} (temperature $0.2$, max tokens 4,096), leveraging its efficiency for synthesizing well-structured, JSON-formatted final trajectories.
\end{itemize}
With this ensemble configuration established, the pipeline executes the following core phases:
\subsection{Phase 1: Label Classification}
The pipeline begins by identifying the label of the issue. The pre-existing repository labels in issue meta-data are first matched with some pre-defined keywords. If the labels do not match with any of the keywords, then the issue title and the problem description are passed to a \textbf{Label Classifier LLM}, which then classifies the thread into a standardized category (e.g., \textit{bug}, \textit{enhancement}, \textit{question}). Crucially, this classification acts as a routing mechanism; it selects a specific \textbf{Field Schema} that dictates the exact semantic fields according to issue type (e.g., \textit{problem\_description}, \textit{reproduction\_context} for \textit{bug} type issue). This is required for downstream trajectory generation.

\subsection{Phase 2A: Contextual Augmentation (Link Content Summarization)}

To address the problem of artifact blindness inherent in text-centric benchmarks, the pipeline explicitly grounds each comment of the issue thread in the corresponding repository artifacts referenced by the developers. In Phase 2A, a URL extraction module scans all issue comments and parses embedded hyperlinks using pattern-based detection. Each detected link is normalized and classified into structured categories before retrieval. The resulting artifacts are then processed through format-specific handlers that transform heterogeneous external resources into a unified textual representation suitable for LLM analysis.

\emph{GitHub-native artifacts} constitute the primary category. Commit URLs are resolved through the GitHub API to retrieve commit metadata and the associated diff, including commit message, modified file paths, and patch hunks. Blob URLs referencing source files are dereferenced to their raw file content; when line anchors are present, the system extracts a contextual window around the highlighted range to preserve surrounding code structure. Pull request and issue links are resolved to their corresponding threads, retrieving titles, descriptions, and discussion content. Fine-grained anchors, including issue comments, review comments, and pull request review nodes, are resolved directly to the referenced comment object rather than the full thread, enabling precise context recovery when developers cite a specific diagnostic statement.

\emph{External web resources} are processed through domain-aware retrieval strategies. Reddit links are fetched via the platform’s JSON endpoint to obtain the post metadata, body text, and a bounded subset of top-level comments. Google Drive artifacts, including Docs, Sheets, and Slides, are exported using Google’s document export APIs, converting the content into plain text or tabular representations depending on the document type. For generic web pages such as documentation sites, blog posts, and release notes, the system issues HTTP requests with a browser-like user agent and performs DOM parsing to extract the primary textual content while removing navigation elements, scripts, and layout boilerplate.

\emph{Structured file formats} are handled by specialized parsers to maximize information extraction fidelity. \emph{Image assets} (PNG, JPG, GIF, WebP, BMP), including GitHub CDN links and common hosting services such as Imgur, are downloaded and passed to a vision-capable language model (gpt-5.4-mini) that produces a textual description of the visual content. PDF documents are processed using page-level text extraction, aggregating content across all pages. Modern Microsoft Word files (.docx) are parsed with structure-aware extraction that includes paragraph text and table cell contents, while legacy Word documents (.doc) are converted using lightweight command-line tools. Spreadsheet files (.xlsx/.xls) are also parsed efficiently. Archive files are intentionally not unpacked to avoid arbitrary file expansion.

In addition to hyperlink artifacts, the pipeline extracts both \emph{fenced and inline code snippets} directly from comment bodies. Fenced blocks preserve declared language metadata when available, enabling downstream models to interpret the code in the correct syntactic context. Inline code segments are captured as supplementary signals and incorporated alongside other extracted artifacts.

All retrieved artifacts are then passed to a \textbf{Link Content Summarizer LLM} that transforms raw artifact content into a concise semantic summary. The summarizer is prompted to produce a concise description emphasizing the artifact’s relevance to the issue’s diagnosis, reproduction, or resolution. Each processed URL is stored in a centralized \textbf{link cache} mapping normalized URLs to their summarized representation. This cache eliminates redundant network retrieval and ensures that multiple references to the same artifact across comments resolve to a consistent context representation. Downstream phases of the pipeline therefore operate over a unified artifact-aware context layer, enabling reasoning over both conversational signals and the underlying technical resources referenced during the issue discussion.

\subsection{Phase 2B: Comment Analysis}
Instead of processing the entire discussion thread as a monolithic text dump, Phase 2B evaluates the discourse comment-by-comment. Using the active \textit{Field Schema} and the enriched \textit{Link Cache}, a \textbf{Field Bucket Classifier LLM} analyzes each comment body. The model identifies highly relevant reasoning signals and extracts them as targeted excerpts. These excerpts are then algorithmically routed into \textbf{Field Buckets} corresponding to the schema fields. To preserve social consensus signals, each routed excerpt retains its attribution metadata, including the \textit{comment\_id}, the \textit{author's name}, and the community \textit{reactions} score.

\subsection{Phase 3: Trajectory Synthesis}
In the final phase, the fragmented, structured evidence is synthesized into a cohesive narrative. The populated \textit{Field Buckets} are passed to a \textbf{Field Summarizer LLM}. This model weighs the collected excerpts, giving precedence to highly-reacted comments and maintainer insights, to generate a paragraph-level summary for each field key. The output is the \textbf{Final Trajectory}, a cleanly formatted JSON object where the community's distributed reasoning is explicitly mapped to the dynamically selected, label-guided fields (e.g., \textit{"root\_cause\_analysis": "<summary>"}).

\section{Dataset Quality Assessment}
Ensuring the high fidelity, technical accuracy, and structural consistency of the extracted reasoning trajectories is critical for their downstream utility in agentic software engineering tasks. Given the complexity and length of real-world GitHub issue threads, manual verification of every generated trajectory is prohibitively expensive. Therefore, we adopted a hybrid evaluation strategy combining manual human inspection with comprehensive automated evaluation via a state-of-the-art LLM.

\paragraph{\textit{Human Evaluation.}}
To establish a baseline for extraction quality and ensure that the pipeline's outputs genuinely reflect the nuances of the original developer discussions, we conducted a preliminary human evaluation on a random sample of issues from the dataset. Annotators with software engineering experience manually reviewed the generated trajectories alongside the original GitHub threads and any associated external artifacts. This manual review confirmed the pipeline's robustness, particularly validating the effectiveness of the external \emph{Link Cache} in grounding the trajectories in actual repository state rather than generic LLM assumptions.

\paragraph{\textit{LLM-as-a-Judge Evaluation (GPT 5.4).}}
To scale the quality assessment across the entirety of the \textbf{SWE-MIMIC-Bench} dataset, we utilized an LLM-as-a-Judge methodology powered by ChatGPT Pro (GPT-5.4 class). We evaluated all generated trajectories across the three dataset variants (SWE-Bench-Multilingual, SWE-Bench-Verified, and SWE-Bench-Pro). The LLM judge was prompted to evaluate each trajectory based on five strict criteria: \textit{Field Coverage}, \textit{Factual Accuracy}, \textit{Technical Depth}, \textit{Structural Faithfulness}, and \textit{Conciseness \& Clarity}. Based on the aggregated scores, the judge assigned a final categorical verdict to each trajectory: \textbf{Excellent}, \textbf{Good}, \textbf{Acceptable}, \textbf{Poor} or \textbf{Inadequate}.
The results of this automated evaluation are summarized in Table \ref{tab:evaluation_verdicts}.

\paragraph{Evaluation Insights.}
The automated evaluation from ChatGPT Pro (GPT 5.4) reveals that the overwhelming majority of the generated trajectories maintain a high standard of quality. Across all three datasets, more than 85\% of the generated trajectories were rated as \textit{Acceptable} or higher. The \textbf{Verified} dataset achieved the highest overall quality scores, with 54\% of its trajectories rated as \textit{Excellent} or \textit{Good}, and only 2.6\% rated as \textit{Poor}. Conversely, the \textbf{Pro} dataset, which features significantly longer and more technically convoluted discussion threads, proved to be the most challenging for the extraction pipeline, resulting in an \textit{Acceptable} rating for the majority (59.3\%) and a slightly elevated \textit{Poor} rate (14.1\%). This disparity highlights the inherent difficulty of tracking multi-party, long-horizon feature design discussions compared to isolated bug fixes, yet the pipeline successfully navigated the extraction without severe degradation.

\begin{table}[t]
\centering
\footnotesize
\setlength{\tabcolsep}{3pt}
\renewcommand{\arraystretch}{1.25}

\caption{Distribution of trajectory quality verdicts assigned by the GPT 5.4 evaluator across the three SWE-Bench dataset variants and their overall aggregation.}
\label{tab:evaluation_verdicts}

\begin{tabular}{lrrrrrrrr}
\hline
& \multicolumn{2}{c}{\textbf{\makecell{Multi\\lingual}}} 
& \multicolumn{2}{c}{\textbf{Verified}} 
& \multicolumn{2}{c}{\textbf{Pro}} 
& \multicolumn{2}{c}{\textbf{All}} \\
\cline{2-3} \cline{4-5} \cline{6-7} \cline{8-9}
\textbf{Verdict} & C & \% & C & \% & C & \% & C & \% \\
\hline
Excellent (E)  &   2 &  0.7 &  19 &  7.0 &   0 &  0.0 &  21 &  2.6 \\
Good (G)       & 108 & 38.3 & 127 & 47.0 &  66 & 26.6 & 301 & 37.6 \\
Acceptable (A) & 148 & 52.5 & 117 & 43.3 & 147 & 59.3 & 412 & 51.5 \\
Poor (P)       &  24 &  8.5 &   7 &  2.6 &  35 & 14.1 &  66 &  8.3 \\
\hline
\textbf{Total} & \textbf{282} & \textbf{100} & \textbf{270} & \textbf{100} & \textbf{248} & \textbf{100} & \textbf{800} & \textbf{100} \\
\hline
\textbf{Total Approved (E+G+A)} & \textbf{258} & \textbf{91.5} & \textbf{263} & \textbf{97.3} & \textbf{213} & \textbf{85.9} & \textbf{734} & \textbf{91.7} \\
\hline
\end{tabular}
\end{table}

\section{Limitations}
While the developed pipeline robustly extracts reasoning trajectories, our methodology has two primary limitations regarding external context retrieval. First, our contextual augmentation in phase 2A relies on targeted regular expressions to fetch common artifacts (e.g., GitHub links, Google Workspace documents etc). However, some unhandled external references may be bypassed, leaving some non-standard, obfuscated context uncaptured. Second, summarizing exceptionally large codebase artifacts, such as monolithic pull requests or extensive commit diffs, introduces inherent scalability challenges. The aggressive text compression required to fit these artifacts within LLM context windows, risks obscuring the highly localized, critical lines of code that actually resolve an issue. Consequently, the technical fidelity of the final trajectory may degrade for tasks involving massive, cross-component refactoring.
\section{Conclusion}
In this work, we introduced \textbf{SWE-MIMIC-Bench}, a benchmark generated via a multi-LLM pipeline designed to extract structured, label-guided trajectories from unstructured GitHub issue discussions. Unlike existing benchmarks that emphasize end-task performance, our approach focuses on modeling the process of reasoning embedded within community-driven discourse. By decomposing issue threads into semantically meaningful fields, and synthesizing them into coherent trajectories, our framework captures the temporal and cognitive evolution of problem-solving in real-world software engineering settings.

Through extensive evaluation on 734 issues from the SWE-Bench Pro, Multilingual and Verified variants, we demonstrate that label-guided trajectories significantly reduce the information overhead required to understand complex discussions while preserving technical fidelity. The results further indicate that most generated trajectories achieve acceptable-to-good quality, validating the effectiveness of our multi-stage pipeline in producing reliable and interpretable representations.

More broadly, SWE-MIMIC-Bench highlights the importance of moving beyond static input–output formulations toward process-centric representations in software engineering research. By exposing structured reasoning pathways, our work provides a foundation for training and evaluating agents that can better emulate expert developers, assist researchers and developers of the software engineering community, and support reinforcement learning from human-like reasoning traces. Future work may explore integrating trajectory supervision into agentic repair systems, expanding label schemas to cover more diverse issue types, and incorporating human-in-the-loop validation to further improve robustness and generalizability.
\label{sec:related}

\section{Availability and Reproducibility}
All data, scripts and detailed documentation are shared in this GitHub repository: 
https://github.com/Geek-a-Byte/SWE-MIMIC-BENCH
\cite{swemimicbench_github_2026}

\bibliographystyle{ACM-Reference-Format}
\bibliography{main}

\end{document}